%Paper: 9111018
%From: <atina!iafe!carmen@uunet.UU.NET>
%Date: Fri, 8 Nov 91 10:27:28 ARG

%%%(Table 1 appended at the end must be processed separately)%%%

\input jnl
\oneandahalfspace
\tolerance=5000
%\magnification=1200
\def\si{\sigma}
\def\La{\Lambda}
\def\inv{\cal N}
\def\sua{\widehat{SU}(N)}
\footnote{}{\singlespace $^*$\ Work partially supported by
CONICET of Argentina, by the Commission of the European Communities
under contract No. C 11-0540-M(TT), and by Fundaci\'on Antorchas.}
\title N=2 COSET COMPACTIFICATIONS WITH NON-DIAGONAL INVARIANTS
\author G. Aldazabal$^{1}$, I. Allekotte$^{1}$, A. Font$^{2}$
and C. N\'u\~nez$^{3,*}$
\affil $^1$Centro At\'omico Bariloche, 8400 Bariloche,
Comisi\'on Nacional de Energ\'{\i}a At\'omica,
Consejo Nacional de Investigaciones Cient\'{\i}ficas y T\'ecnicas,
and Instituto Balseiro, Universidad Nacional de Cuyo - Argentina
\affil $^2$Departamento  de F\'{\i}sica, Facultad de Ciencias,
        Universidad Central de Venezuela,
        A.P.20513, Caracas 1020-A - Venezuela
\affil $^3$?%\iafedir

\abstract

We consider 4-dimensional string models obtained  by tensoring
N=2 coset theories with non-diagonal modular invariants.
We present results from  a  systematic \nobreak analysis including
moddings by discrete symmetries.

\endtitlepage

\leftline{\bf 1. Introduction}

The almost uniqueness of 10-dimensional $SO(32)$ or $ E_8\times  E_8$
heterotic  string theory, once thought to be a desirable  property
for a ``theory of everything", has  now been swept out by  the
rapidly growing production of 4-dimensional string theories.
Awkward as this may seem however, all these chiral theories in
lower dimensions
share with the original one all the  other features
that started the excitement: anomaly freedom, potential for realistic
phenomenology and the promise of a finite and consistent
theory of quantum gravity.

With  all these theories at our disposal the idea of thoroughly
studying each of  these models seems discouraging. It appears
more reasonable instead to try  to understand general properties
of the different constructions, to find  relations among  the
diverse models, in order to gain a better comprehension of the
theory. The present paper aims at this general goal.

One very interesting class of string compactifications was
introduced by Gepner \refto{Gepner87}, who showed how
to obtain (2,2) models by taking tensor products of
$N=2$ minimal models for the internal degrees of freedom. The
spectrum of these compactifications has been
constructed when $A,D$ or $E$-type
invariants of the $SU(2)$ affine algebra are used$^{[
\cite{Gepner87},\cite{Lutken},\cite{LS1},\cite{Fuchs}]}$.
The $N=2$ minimal models of the Gepner construction can be replaced
by the $N=2$ coset theories found by Kazama and Suzuki\refto{Kazama},
thus increasing the possibilities for
the internal part of the compactified  string  theories.
Models obtained  by tensoring the simplest $CP_m$ cosets
with diagonal invariants
(referred to as $CP_m$ diagonal models in the following)
have been analyzed in Ref.\[{Font}].
Other cosets have been investigated in Refs.
[{\cite{Bailin},\cite{But1},\cite{But2},\cite{LS2},\cite{Sche}}].

 It is then natural
to extend the analysis of models based on $N=2$ cosets to
include non-diagonal invariants.
Such computation is
not only useful for the aim of classifying all string vacua but
also as a systematic search  of promising phenomenological models.
 Following this direction
we  have implemented the computation of the number
of generations in $CP_m$ models
with different kinds of invariants.

Although a complete classification only
exists in the $SU(2)$ case \refto{Capelli}, many positive, off-diagonal
invariants can be constructed for other algebras
$^{[\cite{Ber},\cite{ALZ},\cite{su3},\cite{SY1},\cite{SY2},
\cite{dual},\cite{Ver}]}$.
These include series invariants
associated with automorphisms of the affine algebra,
invariants obtained by conformal embeddings and invariants
obtained by acting with an automorphism of the fusion rules of
the extended algebra. In the appendix we have
explicitly listed  various types of $SU(N)$ invariants,
including some that, to our knowledge,
have not appeared previously in the literature.

This article is organized as follows. In section 2 we review
the basic notions of the Gepner construction
as applied to $CP_m$ cosets. In section 3 results are
listed and discussed. In section 4 we include  moddings
by discrete symmetries. Conclusions are presented in section 5.
Non-diagonal invariants for the $SU(N)$ affine algebra
are described in the appendix.
\bigskip

\leftline{\bf 2.  $N=2$ coset theories and 4-$d$ strings}

The $N=2$ coset theories of Kazama and Suzuki\refto{Kazama} are found by
applying the usual GKO construction to a coset $G \times SO(2D)/H$
where $D={1\over 2}\ dim \ (G/H)$. When $G/H$ is a hermitian
symmetric space this construction does provide a
realization of the $N=2$ superconformal algebra (SCA).
In this note we will only deal with the so-called projective
cosets $ G/H=CP_m={SU(m+1)/SU(m)\times U(1)}$ with
central charge given by $c= 3km/(k+m+1)$. In particular,
the $N=2$ minimal models are obtained as $CP_1$ cosets.
A $CP_m$ model at level $k$ is denoted by $(m,k)$ for $m>1$
and by $k$ for $m=1$.

States of the SCA are denoted by
$|\Lambda, \lambda, \tilde \Lambda\rangle$
where $| \Lambda \rangle$
and $|\tilde \Lambda \rangle$ are respectively highest weights
of $G$ (level $k$) and $SO(2D)$ (level 1) restricted by unitarity.
The weight $|\lambda \rangle$ of the $H$ algebra (level $k+1$)
 is found by
decomposing the $G \times SO(2D)$ representations into
irreducible representations of $H$.
For $CP_m$ cosets it is useful to split $\lambda$ into an $SU(m)$
weight $\hat \lambda$ and a $U(1)$ charge $q$ as
$$\lambda = \hat \lambda + {q\over m}w_m \eqno(lt) $$
where $w_m$ is the $m^{th}$ fundamental weight of $SU(m+1)$.
By definition, $\hat \lambda \cdot w_m=0$ and
$q$ is defined mod $m(m+1)(k+m+1)$.
\par
The conformal dimension of a
$|\Lambda, \hat \lambda, q,  \tilde \Lambda \rangle$
state is given by
$$h = {1 \over {2(k+m+1)}} \lbrack C(\Lambda) - C(\hat \lambda)
- {{q^2}\over {m(m+1)}} \rbrack + {1\over 2} \tilde \Lambda^2 + n
\eqno(cd) $$
where $C(W)$ is the quadratic Casimir of the representation
with highest weight $W$ and $n$ is the level at which the
state appears.
The $U(1)$ superconformal charge of the primary
states is given by
$$Q = -{q\over {k+m+1}} + \sum_{l=1}^{D}
\tilde \Lambda^l \eqno(sq)$$
The chiral states of the SCA are those primaries with
$Q=-2h$. From the above eqs. we see that for $\tilde \Lambda =0$
(and $n=0$) they must have $\Lambda =\lambda$.
Likewise, there are anti-chiral
states with $Q=2h$, whose weights are conjugate to those of
the chiral states.
\par

Kazama and Suzuki\refto{Kazama} showed how to generalize the Gepner
construction to obtain modular
invariant, space-time supersymmetric 4-$d$ heterotic strings
in which the internal $c=9$, $N=2$ theory is given by a tensor
product of $N=2$ coset theories.
For a given sector, say the right one,
we denote the internal product of $r$ cosets by
$$\otimes_{i=1}^{r} \ |\Lambda_i, \hat \lambda_i, q_i,
\tilde \Lambda_i \rangle \eqno(ten)$$
The complete state is obtained by appending
the space-time part to \(ten) and
by coupling it to the corresponding expression
for the left sector. We now review briefly the basic
elements of the heterotic construction which imposes
restrictions on the possible quantum numbers of the states.
\par
To describe the generalized GSO projection that guarantees
space-time supersymmetry it is useful to represent the right part of a
given state by a vector
$$V_R = (\tilde \Lambda_0; q_1, \cdots, q_r; \tilde \Lambda_1,
 \cdots,\tilde \Lambda_r) \eqno(vr)$$
where $\tilde \Lambda_0$ is the $SO(2)$ weight that gives
the space-time properties of the state. $V_L$ has a similar
expression but with $\tilde \Lambda_0$ an $SO(10) \subset E_6$
weight that gives the gauge properties of the state.
 We also define the scalar product
$$V \cdot V' = \sum_{i=0}^{r} \ \tilde \Lambda_i
\tilde \Lambda'_i \ - \ \sum_{i=1}^{r} \
{{q_iq'_i}\over {2\eta_i (k_i + m_i + 1)}} \eqno(ps)$$
where $\eta_i=m_i(m_i+1)/2$, and the vectors
$$\eqalign{\beta_0 &= (\bar s; \eta_1,
\cdots, \eta_r;
s_1, \cdots, s_r) \cr  \beta_i &= (v; 0, \cdots, 0; \cdots, v, \cdots)
\ \ \ (v \ \hbox{in the} \ i^{th} \ \hbox{position)} \cr} \eqno(bet)$$
where $s, \bar s$ and $v$  respectively stand for
the highest spinor, anti-spinor
and vector weights of the appropriate $SO$ groups.
\par
The generalized GSO projection consists of imposing
$$\eqalign{
 2\beta_0 \cdot V_R = \hbox{odd integer} \cr
 2 \beta_i \cdot V_R = \hbox{even integer} \cr} \eqno(GSO)$$
Modular invariance forces
the left vector $V_L$ to be related to $V_R$ through
$$\tilde V_L = V_R + n_0\beta_0 + n_i\beta_i \eqno(condlr)$$
where $\tilde V_L$ is obtained by adding a vector
$(v;0, \cdots, 0;0, \cdots,0)$ to $V_L$.
The $SU(m+1)$ and $SU(m)$ left and right weights of
each block are correlated according to an invariant
of these algebras.
In this note we will use $SU(m+1)$
invariants given by series that occur at generic level as well
as exceptional invariants that occur at particular levels.
These types of invariants are described in the appendix.
For $SU(m)$ we will only use the diagonal invariant and postpone the
more general case for future work.
\par
The massless spectrum of the resulting heterotic compactification
consists of the supergravity and Yang-Mills multiplets plus
chiral multiplets transforming as 27, $\overline{27}$ or 1's
under the observable $E_6$. We are mainly interested
in computing the number of 27 and $\overline {27}$'s.
This is most easily done by looking at the scalar component that
transforms as a 10 of $SO(10)$. For the right sector
we then need states with
$h^R_{int} = {1\over 2}$ and $Q^R_{int}= - 1$.
For the left part we need $h^L_{int}= {1\over 2}$ and
$Q^L_{int} = -1$ ($Q^L_{int} = 1$) for 27's
($\overline {27}$'s).
Recall that under $SO(10)\times U(1)$ a 27 decomposes as
27 = $16_{{1\over 2}} + 10_{-1} + 1_2$.
\par
Only chiral states in the sub-theories can produce tensor states
with $h_{int} = {1\over 2}$ and $Q_{int}= - 1$ whereas for
$Q_{int}= 1$ only anti-chiral states contribute. Generations
(anti-generations) are thus obtained by combining right chiral
and left chiral (anti-chiral) states according to the rules explained
above.

To avoid multiple counting of states
it is necessary to take into account possible field identifications
\refto{Gepner89}.
These might appear due to the existence of
a proper external automorphism $\sigma$ acting on the affine
algebras in the coset construction.
Under this automorphism $\Lambda$ transforms as
$$\sigma(\Lambda) = kw_1 + a(\Lambda)\quad \eqno (siglam)$$
where $w_1$ is the 1st fundamental weight of $SU(m+1)$ and $a$
is the Coxeter rotation acting on the fundamental weights as
$$a(w_i) = w_{i+1} - w_1 \eqno(Cox)$$
$\hat \lambda$ transforms in a similar way with $k$ replaced by
$(k+1)$. The $U(1)$ charge transforms as
$$\sigma(q) = q + (k+m+1) \quad \eqno (sigq)$$
The effect of $\sigma$ on $\tilde \Lambda$ is to exchange
the singlet with the vector weight and the spinor with the
anti-spinor weight. The order of $\sigma$ is $m(m+1)$.

This automorphism may act on the left and right sectors independently.
However, if the modular coefficients are invariant under a certain
subgroup of $\sigma$ transformations
$$
{\cal N}_ {\sigma^{\alpha}(\Lambda),\sigma^{\beta}(\Lambda')}
={\cal N}_ {\Lambda, \Lambda'}
\eqno (modcoef)$$
the following field identifications
$$ \eqalign{
|\Lambda, \hat \lambda, q,  \tilde \Lambda \rangle_R &\otimes
|\Lambda', \hat \lambda', q',  \tilde \Lambda' \rangle_L = \cr
&|\si^\alpha(\Lambda), \si^\alpha(\hat \lambda), \si^\alpha(q),
\si^\alpha(\tilde \Lambda) \rangle_R \otimes
|\si^\beta(\Lambda'), \si^\beta(\hat \lambda'), \si^\beta(q'),
\si^\beta(\tilde \Lambda') \rangle_L  \cr}\eqno(ident)$$
 are necessary.
Notice that $\alpha = \beta$ mod $m$ must be required
since we are considering the
diagonal invariant for $SU(m)$.

To clarify the issue of field identifications
let us consider a coset model built up of two theories
coupled through invariants satisfying
 \(modcoef) for some $\alpha_i, \beta_i$.
If a state
$$
[\mid\Lambda_1, \hat \lambda_1, q_1, \tilde \La_1 \rangle
 \mid \Lambda_2, \hat \lambda_2,q_2, \tilde \La_2 \rangle ]_R \otimes
[\mid\Lambda'_1,\hat \lambda'_1,q'_1, \tilde \La'_1 \rangle
 \mid \Lambda'_2, \hat \lambda'_2,q'_2, \tilde \La'_2 \rangle ]_L
\eqno(state)$$
is in the spectrum we know from \(condlr) that ($i=1,2$)
$$
q'_i-q_i =  {n_0(m_i+1)m_i\over 2} \ \hbox{mod } m_i(m_i+1)(k_i+m_i+1)
\eqno (qq)$$
Let us now transform the right (left) sector of the above state \(state)
 by the action of $\sigma^{n_1 \alpha_1}$, $\sigma^{n_2 \alpha_2}$
 ($\sigma^{\bar n_1  \beta_1}$, $\sigma^{\bar n_2 \beta_2}$)
 to obtain a new state
$$\eqalign{
[\si^{n_1 \alpha_1}\mid \La_1,\hat \lambda_1, q_1, \tilde \La_1
\rangle &
\si^{n_2 \alpha_2} \mid\La_2, \hat \lambda_2, q_2, \tilde \La_2
\rangle ]_R  \otimes \cr
&[\si^{\bar n_1 \beta_1}\mid\La'_1,
\hat \lambda'_1,q'_1, \tilde \La'_1\rangle
\si^{\bar n_2 \beta_2}\mid\La'_2,
\hat \lambda'_2,q'_2, \tilde \La'_2 \rangle  ]_L \cr}
\eqno(eqstate)
$$
This new state
will have the same $Q$ and $h$ as the original one and
will be allowed by the modular invariant due to \(modcoef).
Thus \(eqstate) will be in the spectrum
if $\sigma^{n_i \alpha_i}(q'_i) -
\sigma^{\bar n_i \beta_i}(q_i)$
satisfy condition \(qq) with some $n_0'$ instead of $n_0$.
 Comparing both sets of equations and using
$ \sigma^{n}(q_i)= q_i + n(k_i+m_i+1)$
with $ n< m_i(m_i+1)$
leads to the conditions
$$(n_i \alpha_i - \bar n_i \beta_i)( k_i+m_i+ 1) =
  (n'_0-n_0){{m_i(m_i + 1)} \over 2}
  \ \hbox{mod } m_i(m_i+1)(k_i+m_i+1) \eqno(cond1)$$
which in turn imply
$$
({{n_1 \alpha_1 - \bar n_1 \beta_1}\over m_1(m_1+1)} + t_1)(k_1+m_1+1)=
({{n_2 \alpha_2 - \bar n_2 \beta_2}\over m_2(m_2+1)}
+ t_2)(k_2+m_2+1) \eqno(cond2)$$
with $t_1,t_2$ integers.

Hence, if conditions \(modcoef), \(cond1)
and \(cond2) are met, state \(eqstate) is also allowed.
As these conditions do not depend on the particular
 state we started with,
states \(state) and \(eqstate) should be identified in order
to avoid a multiple counting. Or, equivalently, the partition function
must carry a factor of ${1 \over M}$ \refto{Gepner89}, where $M$, the
multiplicity of all the states, can be deduced
from the above formulae.
The extension of
the previous result for the case of $r$ theories is achieved by considering
theories in pairs.
\par
All but one of the invariants considered by us (namely, the series $G$
invariant for $SU(4)$, $k$ odd) verify the relation
$$
{\cal N}_{\sigma(\Lambda),\sigma(\Lambda')} =
 {\cal N}_{\Lambda,\Lambda'}
\eqno(invdiag)
$$
In this case, the field identification \(ident)
with $\alpha=\beta=1$ is required.
If the invariant possesses further symmetries, such as
$$
{\cal N}_{\sigma(\Lambda),\Lambda'} = {\cal N}_{\Lambda,\Lambda'}
\eqno(invd)
$$
new identifications have to be considered.

As an example, consider the tensoring of two Gepner models.
For $m_i=1$ condition \(cond1) can  always be fulfilled.
$SU(2)$ invariants satisfying \(invd)
are $D_{even}$, $E_7$ and $E_8$, with $k_i=0$ mod $4$.
Using these invariants for both blocks gives
a multiplicity factor of $8$, as can be seen
by inspecting the values of $n_i$,
$\bar n_i$ for which equation \(cond2) can be satisfied.
\bigskip

\leftline{\bf 3. Results}

We have computed the number of $E_6$ generations $N_{27}$
and antigenerations $N_{\overline{27}}$\ for tensor
products of $CP_m$ coset models having at least
one non-minimal block and admitting at least one
non-diagonal invariant of the types described in the appendix.
These numbers have been obtained
by explicitly constructing massless states in the tensor theory
as explained in section 2. As a check on our algorithm we
analyzed tensor products with $c=6$ and verified that they
correspond either to the $K3$ manifold or to the torus.
Results for tensor models with $c=9$ are
presented in Table 1 (the numbering of models follows
that of Ref.\[Font]).

The net number of generations,
$N_{gen}=|N_{27}-N_{\overline{27}}|$,
varies between 0 and 360.
Similarly as in Refs.
[\cite{Lutken},\cite{Fuchs},\cite{Font},\cite{LS2},\cite{schya}]
most values of $N_{gen}$
are multiple of 12, 18 or 8. A few other models,
which will be described in the next section, present a lower
common divisor.
Models with 0 generation are also numerous,
they add up to 72, out of 357 models listed. Those with
$N_{27}=N_{\overline{27}}=21$ actually correspond to
compactifications on the manifold $K_3 \times T$ as observed
in Ref.\[Lutken].
\par
For minimal models with $A$-$D$-$E$ invariants and for $CP_m$ cosets
with diagonal invariants the superpotential $W$ which defines the
corresponding Calabi-Yau manifold is known$^{[\cite{VW},\cite{Mar},
\cite{GVW},\cite{GepnerW},\cite{LVW}]}$.
Given the polynomial $W$ the net number of generations can be
found using a formula derived by Vafa \refto{Vafa}.
For the models involving diagonal invariants in the non-minimal blocks
we have computed $N_{gen}$ from the Landau-Ginzburg (LG) formulation
and found complete agreement with the explicit calculations.

The LG formulation of $N=2$ models also allows to demonstrate relations
among theories.
Some known $^{[\cite{GVW},\cite{Schwarz}, \cite{Font}]}$
equivalences are
$10_E=1_A 2_A$, $28_E=1_A 3_A$, $4_D=1_A 1_A$, $(2,2) = 8_D$
and $(m,k)_A=(m-1,k+1)_A (1,{k-m+1\over m})_A$.
{}From the form of the superpotential for diagonal $CP_m$ models
we can obtain the following additional equivalences:
$(3,4)_A=6_A 6_D$, $(3,5)_A=7_A 16_E$, and
$(4,7)_A=4_A 10_A 10_E$.
We have checked these relations
explicitly by computing their respective spectra.
To simplify the presentation, Table 1 does not include
models related by the above equivalences.
For non-diagonal $CP_m$ models
further empirical relations can be established among theories,
e.g. $(2,9)_C = (3,4)_C$, $(4,5)_E=(4,5)_C$ and $(3,k$ odd$)_G=(3,k$ odd$)_A$.

Given the relation between $CP_1$ minimal theories with $A$-$D$-$E$
invariants and modality zero
singularities$^{\lbrack \cite{Mar}, \cite{VW} \rbrack}$
it is natural to ask whether similar results apply to the non-diagonal
$CP_m$ theories that we have analyzed. More importantly, one
would like to know whether these theories admit a LG description.
We have only been able to answer this question in the affirmative
in a few isolated cases. For example, we find
$(3,2)_{D_2} \equiv 1_A 4_D$, $W=x^3 + y^3 + yz^2$ and
$(5,2)_G \equiv 2_A 6_D$, $W= x^4 + y^4 + yz^2$. In other cases
the question has a negative answer. For instance, the $(2,3)_D$
theory has only one chiral field with $Q \geq 1$ and thus the
putative LG polynomial would have modality one. But none
of the modality one singularities that have been completely
classified\refto{Arnold} has the right properties to match
those of the $(2,3)_D$ theory. For generic theories we claim,
without rigorous proof,
that a quasihomogeneous superpotential cannot be defined.

In a recent work Buturovi\'c\refto{But2} has
developed a method to compute the number of
families for models that do not admit a LG formulation
but whose Poincar\'e polynomial ${\cal P}$ is known. For most of our
non-diagonal $CP_m$ models we have derived the corresponding
${\cal P}$ and checked that the results for $N_{gen}$ using this method
agree with the explicit calculations of the massless states.
In particular, the equivalences $(2,9)_C = (3,4)_C$ and
$(4,5)_E=(4,5)_C$ mentioned above, follow from their corresponding
${\cal P}$'s.
\bigskip

\leftline{ \bf 4. Moddings by discrete symmetries}

The values of $N_{gen}$
displayed in Table $1$ are rather large
and phenomenologically not viable.
However, in general, $CP_m$ models possess
discrete symmetries that upon modding lead to
a reduced number of families.
For instance, the only known 3-generation model is obtained from
the tensor product $1(16_E)^3$ by modding by a $Z_3 \times Z_3$
symmetry \refto{G3}. With this motivation in mind
we have included moddings in our analysis.
\par
Discrete symmetries and moddings of tensor products of $CP_m$
cosets have been studied in detail in Ref. \[{FIQS}]. The
symmetry group of a tensor product model is given by
${\cal G} = {\cal A} \times {\cal P}$ where
${\cal A} = Z_{k_1 + m_1 + 1} \times
 \cdots \times  Z_{k_r + m_r + 1}$ and ${\cal P}$ is the permutation
group of identical blocks. A new model can be obtained by dividing
by any subgroup of ${\cal G}$. We will mainly focus on subgroups of
${\cal A}$ and now describe the simplest case of modding by a
$Z_M$ subgroup.
\par
$Z_M$ is generated by the modding vector
$$\Gamma = (0; \eta_1 \gamma_1, \cdots, \eta_r \gamma_r;0) \eqno(uno)$$
The order $M$ is the least integer such that
$M(\gamma_1, \cdots , \gamma_r) = 0 \hbox{ mod } (k_1+m_1+1,
\cdots , k_r + m_r + 1)$. Dividing by $Z_M$ introduces twisted sectors
in which
$$\tilde V_L = V_R + n_0\beta_0 + n_i\beta_i + 2x\Gamma \eqno(dos)$$
where $x=0,\cdots , M-1$. Furthermore, the projection
$$ \Gamma \cdot ( \tilde V_L + V_R) = -
\sum_{i=1}^{r} \ { { \gamma_i(\bar q_i + q_i)}\over {2(k_i+m_i+1)} }
= \hbox{integer} \eqno(tres)$$
selects the allowed states in each sector.
\par
To preserve supersymmetry $\Gamma$ must verify the condition
$$ 2\beta_0 \cdot \Gamma = -
\sum_{i=1}^{r} \ { { \eta_i \gamma_i}\over {(k_i+m_i+1)} }
= \hbox{integer} \eqno(cuatro)$$
There is a finite number of nonequivalent $\Gamma$'s for a given
tensor model. Two moddings $\Gamma_1$ and
$\Gamma_2$ are equivalent if
$$ \Gamma_1 = \Gamma_2 + 2n(0;\eta_1, \cdots , \eta_r;0) \eqno(cinco)$$
since $ 2(0;\eta_1, \cdots , \eta_r;0)$ acts trivially on the spectrum.
\par
We have systematically included all nonequivalent single moddings
in combination with non-diagonal invariants. A complete list of results
is available on request.
We found four models with 4 generations, namely
$(3,6)_C3_A3_A$, $(3,3)_A12_A12_A$, $(2,9)_E3_A18_{A,F}$ and
$(2,9)_E6_A6_{A,F}$.
Models with 6 generations are $(2,9)_D(2,9)_D$, $(2,24)_{A,D}1_A16_{A,D}$
and $(2,9)_{D,E}4_A10_{A,F}$. The remaining models still have
8 or more families.

A generic feature of
the results is the fact that for a given model the number of
generations obtained through different moddings is always a multiple
of a fixed integer. Moreover, replacing diagonal by series
invariants does not alter this integer.
On the contrary, exceptional invariants may give models
with a smaller common factor.
As an example consider model $60: (2,9)(2,9)$.
For diagonal and series
invariants $N_{gen}$ is always a multiple of $6$.
Model $60EE$ (with the
exceptional $E(2,9)$ invariant) with all possible moddings has a common
divisor $D=2$. Model $190$ also possesses $D=2$.
A model with $D=3$ is $(2,24)_{A,D}1_A16_E$.
Some models with a fundamental factor
of $4$ and $6$  also arise. They are $39$, $43$, $45$, $50$, $153$,
$170$, $176$  and
$9$, $60$, $129$, $144$, $172$ respectively.
This sort of quantization property has also been noticed in
Refs. $\lbrack \cite{FIQS}, \cite{schya} \rbrack$.

It is also interesting to notice that in most of the cases replacing a
diagonal by a series invariant for one block gives the same
result as a modding  with the diagonal invariant.
\bigskip

\leftline{\bf 5. Conclusions}
As a step forward in the program of a systematic
analysis of possible (2,2) string vacua,
we have extended previous results by considering $CP_m$ coset
models with a rather large class of non-diagonal invariants.

A global inspection of Table $1$
shows that the results for the net number
of generations do not differ appreciably from the ones obtained with
diagonal invariants or even from minimal model constructions.
This could be
surprising in the sense that, in most of the cases,
non-diagonal invariants seem to impose
severe restrictions on the possible left-right pairings.
This is however compensated by the need of considering
$\sigma$-transformed states, as explained in Section 2.

In order to lower the number of generations,
we have modded out by discrete symmetries performing
all possible single moddings combined with off-diagonal invariants.
We found that for a given model $N_{gen}$ is always multiple
of a fixed fundamental number. This is an interesting and
useful result since it indicates which models could potentially
have 3 families upon further modding by $Z_M$ or permutation
symmetries.

As a byproduct of our analysis we found some isomorphisms,
e.g. $(4,5)_C=(4,5)_E$, between $N=2$ $CP_m$ theories.
Such equivalences are interesting since they could encode
intrinsic properties of the theories.

Directions for future work include
the analysis of $CP_m$ cosets with non-diagonal
couplings for the $SU(m)$ algebra and the systematic
study of other Hermitian symmetric spaces with different
types of modular invariants.

\bigskip\bigskip
\leftline{\bf Acknowledgements}
A.F. wishes to thank: 1) the CAB and the IAFE for hospitality
as well as the Universidad de Buenos Aires
for financial aid during the time that this
work was started; 2) the Centro Cient\'{\i}fico IBM-Venezuela
for the continuous use of its facilities.
C.N. thanks the CERN Theory Division and ICTP for hospitality
and financial support while some of this work was carried out.
\vfill\eject

\leftline{\bf APPENDIX. Modular Invariants for $\sua$}

Although a complete classification of $\sua_k$ modular invariants does
not exist, many positive invariants are
known$^{[\cite{Ber},\cite{ALZ},\cite{su3},\cite{SY1},\cite{SY2},
\cite{dual},\cite{Ver}]}$.
They include type $D$ invariants that occur
for generic $k$ and type $E$ invariants that only occur at particular
values of $k$. $D$ invariants that can be constructed systematically
by using properties of automorphisms of the extended Dynkin
diagram$^{[\cite{Ber},\cite{ALZ}]}$ or equivalently by the method
of simple currents$^{[\cite{SY1},\cite{SY2}]}$ are
described in section A.2. $E$ invariants that can be constructed
using conformal embeddings are described in section A.3. Other
exceptional invariants are described in section A.4.

\bigskip
\leftline{\bf A.1 Definitions and Notation}

The unitary irreducible representations of $\sua_k$ are
labelled by a highest weight
$$ \La = \sum_{i=1}^{N-1} \ n_iw_i \eqno(hws)$$
where $w_i$ are the $SU(N)$ fundamental weights ($w_i \equiv
w_{i+N}, w_N \equiv 0$) and $n_i \geq 0$ are the Dynkin labels.
Unitarity imposes the constraint
$$\sum_{i=1}^{N-1} \ n_i  \ \leq k \eqno(uni)$$
The conformal dimension of the primary state labelled by $\La$
is given by
$$h(\La) = { {\La\cdot(\La + 2\rho)}\over {2(k+N)}} \eqno(pc)$$
where $\rho$ is the sum of the fundamental weights. For
future reference we also introduce the
 $N^{th}$-ality of $\La$ defined as
$$ t(\La) = \sum_{i=1}^{N-1} \ in_i \eqno(tri)$$
$t$ is defined mod $N$.
\par
The affine algebra has a group of automorphisms
 with elements $\si^r , r=1, \cdots, N$.
 $\si^r$ acts on $\La$ as
$$  \si^r(\La) = kw_r + a^r(\La)\eqno(aut)$$
where $a$ is the Coxeter rotation belonging to the Weyl group
whose action on the fundamental weights is given by
$$a^r(w_i) = w_{i+r} - w_r\eqno(act)$$
$\si^r$ has order $N(r)$ where $N(r)$ is the least integer
such that $rN(r) = 0  \ \hbox{mod} \ N$.

The $\sua$ specialized character of the representation with weight
$\La$ is given by \refto{Kac}
$$\chi_{\La}(\tau) = {1\over {\eta^{N^2-1}}} \sum_{y \in M}
\left [ \prod_{\alpha^+} { {((k+N) + \La + \rho)\cdot \alpha}\over
{\rho \cdot \alpha} }\right ] \exp [ i\pi(k+N)\tau
(y + {{\La + \rho}\over {k+N}})^2 ] \eqno(wk)$$
where $\eta$ is the Dedekind function and $M$ is the root lattice.
We will often denote characters by $\chi_{n_1n_2 \cdots}$ where
$n_i$ are the Dynkin labels.
A modular invariant is a combination of characters
$$Z(\tau) = \sum_{\La,\La'} {\cal N}_{\La,\La'}
 \chi^{}_{\La}(\tau) \chi^*_{\La'}(\tau)
\eqno(pfn)$$
that remains invariant under the transformations generated by
$\tau \rightarrow \tau + 1$ and $\tau \rightarrow -1/\tau$.
The interpretation of $Z(\tau)$ as a partition function requires
that the ${\cal N}_{\La,\La'}$ be integers
and that ${\cal N}_{0,0}=1$.
In the next sections we will discuss several types of
modular invariant matrices ${\cal N}_{\La,\La'}$
\bigskip

\leftline { \bf A.2  Automorphism invariants }

In this section we review the type of invariants that can
be associated with automorphisms of the affine algebra.
We will first discuss the series of invariants
constructed by Bernard\refto{Ber} and in the end mention briefly
other series that can be built by other
methods~$^{[\cite{ALZ},\cite{SY1}]}$.

In the Bernard construction
an invariant $\inv$ can be associated to the
automorphism $\si^r$ whenever the quantity
$$j(r) = {1\over 2} N(r) (k+N) {{r(N-r)}\over N} \eqno(jd)$$
satisfies the condition
$$2j(r) \equiv  0 \ \hbox{mod 2 and mod}\ N(r) \eqno(jc)$$
Then there exists a modular invariant ${\cal N}^{(r)}$ given by
$${\cal N}^{(r)}_{\La',\La} = \cases {
 \sum_{p=1}^{N(r)} \
(-1)^{p\lbrack r(N-1) + {{2j(r)}\over {N(r)}} \rbrack } \
\delta_{\La', \si^{rp}(\La)}
&if $t(\La)=j(r) + {{N(N-1)}\over 2}$
mod $N(r)$ \cr\noalign{\medskip}
0 & otherwise \cr}\eqno(binv)$$
We now describe the type of invariants obtained in this way
when $N \leq 7$.
\par
When $N$ is an odd prime we can check that, as expected, all values
of $r$ are equivalent. In this case condition \(jc) is satisfied
for $k=0$ mod $N$ and eq. \(binv) yields directly a positive invariant
denoted $D$. This $D$-type invariant turns out to be
$$D_{\Lambda',\Lambda} =  \cases {
 \sum_{p=1}^{N} \
\delta_{\La', \si^{p}(\La)}
&if $t(\La)= 0$ mod $N$ \cr\noalign{\medskip}
0 & otherwise \cr}\eqno(di)$$
\par
We now consider $N$ even. In this case
the situation is more elaborate. In general, there
exist invariants when $r$ and $N$ have common factors. For our
purposes we only need consider $r=1,2,N/2$. When $r=1$ condition \(jc)
again requires $k=0$ mod $N$. If $k=0$ mod $2N$ eq.\(binv) gives a
positive invariant of the above type $D$. If $k=N$ mod $2N$ eq.
\(binv) does not give a positive invariant but positivity is restored
in the combination $F = {\cal N}^{(2)} -  {\cal N}^{(1)}$. In this
way we obtain the $F$-type invariant
$$F_{\Lambda',\Lambda} =  \cases {
 \sum_{p=1}^{N/2} \
\delta_{\La', \si^{2p}(\La)}
&if $t(\La)= 0$ mod $N$ \cr\noalign{\medskip}
 \sum_{p=1}^{N/2} \
\delta_{\La', \si^{2p-1}(\La)}
&if $t(\La)={N\over 2}$  mod $N$ \cr} \eqno(finv)$$
By itself ${\cal N}^{(2)}$ gives a positive invariant
provided $k=0$ mod $N/2$. In fact,
${\cal N}^{(r)}$ gives a positive invariant when
$r$ even divides $N$ and $k=0$ mod $N/r$.
This $D_r$-type invariant has the form
$$[D_r]_{\Lambda',\Lambda} =  \cases {
 \sum_{p=1}^{N/r} \
\delta_{\La', \si^{rp}(\La)}
&if $t(\La)= 0$ mod $N/r$ \cr\noalign{\medskip}
0 & otherwise \cr}\eqno(dr)$$
Finally we consider $r=N/2$. For $N=0$ mod 8 and any $k$, or
for $N=4$ mod 8 and $k=0$ mod 2, or for $N=2$ mod 4 and $k=0$ mod 4,
condition \(jc)
is verified and eq. \(binv)
yields a positive invariant of type $D_{N/2}$.
If $N=4$ mod 8, $k=1$ mod 2 or $N=2$ mod 4, $k=2$ mod 4 condition
\(jc) is also verified but ${\cal N}$ is non positive. Positivity
is restored by subtracting the diagonal invariant. In this
way we obtain the $G$-type invariant
$$G_{\Lambda',\Lambda} =  \cases {
\delta_{\La', \La}
&if $t(\La)= 0$ mod 2 \cr\noalign{\medskip}
\delta_{\La', \si^{N/2}(\La)}
&if $t(\La)=1$  mod 2 \cr} \eqno(gi)$$
Table 2 gives the different invariant types for
$N \leq 7$ according to the above discussion.

There exist other series of invariants associated with
automorphisms (or simple currents) that are not included
in the Bernard construction$^{\cite{ALZ},\cite{SY1}}$.
For instance, for $\widehat {SU}(3)$,
$k \not= 0$ mod $3$, there is an invariant of the form
$$ \sum_{all \ orbits \ of \ \si} |\chi_{\La_0}|^2
+ \chi_{\La_1}\chi^*_{\La_2} + \chi_{\La_2}\chi^*_{\La_1}
\eqno(raros)$$
where $t(\La_0)=0$ mod $3$ and $\La_i = \si^i(\La_0)$.
In our analysis of coset compactifications we did not consider
this type of invariants in order to reduce the number of models.

\bigskip

\leftline{\bf A.3 Conformal Embeddings}

An affine algebra $\hat {\bar G}$ at level $\bar k$ is conformally
embedded into $\hat G$ at level $k$ if $\bar G \subset G$ and their
central charges are equal. All conformal embeddings have been
classified\refto{emb} and shown to require $k=1$. Table 3
displays the embeddings of $\sua$, $N=3, \cdots ,7$ into simple algebras.

When $\hat {\bar G}$ is conformally
embedded into $\hat G$ a character $\chi_{\La}$ of
$\hat  G$ can be written as a finite sum of
characters $\bar \chi_{\bar \La}$ of $\hat {\bar G}$
for which
$$h(\La)= h(\bar \La) \ \hbox{ mod integer} \eqno(hc)$$
To determine the branching rule of $\La$ the first step is
to find the states $\bar \La$ satisfying condition \(hc).
Next, the asymptotic behavior of $\chi_{\La}$ is matched
with that of an appropriate combination of $\chi_{\bar \La}$'s
\refto{Kac}. Finally, the first terms in the expansion of the
characters in powers of $q=\exp(2i\pi\tau)$ are compared.

{} From the above discussion it is clear that modular invariants for
$\hat {\bar G}$ can be computed from invariants of $\hat G$
at level 1. For instance, the $E_6$ and $E_8$ invariants
of $\widehat {SU}(2)$ can be derived in this way\refto{BN}.
Invariants of $\widehat {SU}(3)$ have been obtained in
Ref. \[su3]. Invariants of $\widehat {SU}(4)_4$ and
$\widehat {SU}(5)_5$ were found in Ref. \[SY2].
We have derived those of $\widehat {SU}(4)_{6,8}$,
$\widehat {SU}(5)_7$, $\widehat {SU}(6)_6$ and
$\widehat {SU}(7)_7$ that are relevant to our analysis.
These are all collected below. They are denoted by $C(N-1,k)$
or $\tilde C(N-1,k)$.
$$\eqalign{
C(2,5) = &|\chi_{00} + \chi_{22}|^2 + |\chi_{20} + \chi_{23}|^2
+ |\chi_{02} + \chi_{32}|^2 +\cr
& |\chi_{30} + \chi_{03}|^2
+ |\chi_{12} + \chi_{50}|^2 + |\chi_{21} + \chi_{05}|^2
\cr\noalign{\medskip}
 C(2,9) = &|\chi_{00} + \chi_{90} + \chi_{09} + \chi_{41}
+ \chi_{14} + \chi_{44}|^2
+ 2|\chi_{22} + \chi_{25} + \chi_{52}|^2
\cr\noalign{\medskip}
 C(2,21) = &|\chi_{0,0} + \chi_{21,0} + \chi_{0,21} + \chi_{4,4}
+ \chi_{13,4} + \chi_{4,13} +\cr
&~~~~~~~~ \chi_{10,10} + \chi_{10,1} + \chi_{1,10} +
\chi_{6,6} + \chi_{9,6} + \chi_{6,9}|^2 +\cr
& |\chi_{6,0} + \chi_{0,6} + \chi_{15,0} + \chi_{0,15}
+ \chi_{15,6} + \chi_{6,15} +\cr
&~~~~~~~~ \chi_{7,4} + \chi_{4,7} + \chi_{4,10}+\chi_{10,4}
 + \chi_{10,7} + \chi_{7,10}|^2
\cr\noalign{\medskip}
C(3,4) = &|\chi_{000} + \chi_{012} + \chi_{040} + \chi_{210}|^2
+ |\chi_{004} + \chi_{400} + \chi_{101} + \chi_{121}|^2
+ 4|\chi_{111}|^2
\cr\noalign{\medskip}
C(3,6) = &|\chi_{000} + \chi_{060} + \chi_{202} + \chi_{222}|^2 +
|\chi_{006} + \chi_{600} + \chi_{022} + \chi_{220}|^2  +\cr
&|\chi_{002} + \chi_{240} + \chi_{212}|^2
+|\chi_{200} + \chi_{042} + \chi_{212}|^2 +\cr
&|\chi_{004} + \chi_{420} + \chi_{121}|^2
+|\chi_{400} + \chi_{024} + \chi_{121}|^2 +\cr
&|\chi_{012} + \chi_{230} + \chi_{303}|^2
+|\chi_{032} + \chi_{210} + \chi_{303}|^2 +\cr
&|\chi_{103} + \chi_{321} + \chi_{030}|^2
+|\chi_{301} + \chi_{123} + \chi_{030}|^2
\cr\noalign{\medskip}
C(3,8) =& |\chi_{000} + \chi_{800} + \chi_{080} + \chi_{008}
+ \chi_{121} + \chi_{412} + \chi_{141} + \chi_{214}|^2 +\cr
&  |\chi_{020} + \chi_{602} + \chi_{060} + \chi_{206}
+ \chi_{303} + \chi_{230} + \chi_{323} + \chi_{032}|^2 +\cr
& 2|\chi_{113} + \chi_{311} + \chi_{133} + \chi_{331}|^2
\cr\noalign{\medskip}
C(4,5) =& |\chi_{0000} + \chi_{5000} + \chi_{0500} + \chi_{0050} +
\chi_{0005} + \chi_{0102} + \chi_{2010} + \chi_{0220} +
\chi_{1022} + \chi_{2201}|^2 +\cr
& |\chi_{0013} + \chi_{3100} +
\chi_{0131} + \chi_{1310} + \chi_{1001} + \chi_{1111}|^2 +
8|\chi_{1111}|^2
\cr\noalign{\medskip}}$$
$$\eqalign{
C(4,7) =&
|\chi_{0000} + \chi_{0330} + \chi_{0403} + \chi_{2002}
+ \chi_{2112} + \chi_{3040}|^2 +\cr
& |\chi_{7000} + \chi_{1033} + \chi_{0040} + \chi_{3200} +
\chi_{1211} + \chi_{0304}|^2 +\cr
& |\chi_{0007} + \chi_{3301} + \chi_{4030} + \chi_{0023} +
\chi_{1121} + \chi_{0400}|^2 +\cr
& |\chi_{0700} + \chi_{0103} + \chi_{3004} + \chi_{2320} +
\chi_{2121} + \chi_{0030}|^2 +\cr
& |\chi_{0070} + \chi_{3010} + \chi_{0300} + \chi_{0232} +
\chi_{1212} + \chi_{4003}|^2 +\cr
& |\chi_{0002} + \chi_{0421} + \chi_{2012}
 + \chi_{2130} + \chi_{2203}|^2
+ |\chi_{0312} + \chi_{1240} + \chi_{2000} +
 \chi_{2102} + \chi_{3022}|^2 +\cr
& |\chi_{2500} + \chi_{1004} + \chi_{2220} + \chi_{0121} +
\chi_{3022}|^2 + |\chi_{1210} + \chi_{4001}
+ \chi_{0052} + \chi_{0222} + \chi_{2203}|^2 +\cr
& |\chi_{5000} + \chi_{0042} + \chi_{2201}
 + \chi_{1213} + \chi_{0220}|^2
+ |\chi_{3121} + \chi_{2400} + \chi_{0005}
 + \chi_{1022} + \chi_{0220}|^2 +\cr
& |\chi_{0250} + \chi_{2100} + \chi_{1222}
 + \chi_{3012} + \chi_{0302}|^2
+ |\chi_{2103} + \chi_{0012} + \chi_{0520}
 + \chi_{2221} + \chi_{2030}|^2 +\cr
& |\chi_{0025} + \chi_{4210} + \chi_{0122}
 + \chi_{1301} + \chi_{2030}|^2
+ |\chi_{1031} + \chi_{0124} + \chi_{5200}
 + \chi_{2210} + \chi_{0302}|^2
\cr\noalign{\medskip}
C(5,6) =& |\chi_{00000} + \chi_{01002} + \chi_{20010} + \chi_{00014}
+ \chi_{02020} + \chi_{10112} + \chi_{21101} + \chi_{41000} +\cr
&~~~~\chi_{00230} + \chi_{01032} + \chi_{03200} + \chi_{11211} +
\chi_{23010} + \chi_{00060} + \chi_{01410} + \chi_{06000}|^2 +\cr
&|\chi_{10001} + \chi_{00103} + \chi_{11011} + \chi_{30100} +
\chi_{00006} + \chi_{01121} + \chi_{10023} + \chi_{12110}  +\cr
&~~~~\chi_{20202} + \chi_{32001} + \chi_{60000} + \chi_{00141} +
\chi_{02301} + \chi_{10320} + \chi_{14100} + \chi_{00600}|^2 +\cr
& 16 |\chi_{11111}|^2
\cr\noalign{\medskip}
\tilde C(5,6) =& |\chi_{00000} + \chi_{00200} + \chi_{02020} +
\chi_{04002} + \chi_{20040} + \chi_{00060} + \chi_{06000}|^2 +\cr
& |\chi_{00100} + \chi_{01110} + \chi_{03011} + \chi_{11030} +
\chi_{05001} + \chi_{10050}|^2 +\cr
& |\chi_{01010} + \chi_{02101} + \chi_{10120} + \chi_{01040}
+ \chi_{04010} + \chi_{12021}|^2 +\cr
& |\chi_{02001} + \chi_{10020} + \chi_{00130} + \chi_{03100}
+ \chi_{11111} + \chi_{02031} + \chi_{13020}|^2 +\cr
& |\chi_{00030} + \chi_{03000} + \chi_{11011} + \chi_{01121}
+ \chi_{12110} + \chi_{20202} + \chi_{03030}|^2 +\cr
& |\chi_{01021} + \chi_{12010} + \chi_{20102} + \chi_{02120}
+ \chi_{10212} + \chi_{21201}|^2 +\cr
& |\chi_{02020} + \chi_{10112} + \chi_{21101} + \chi_{30003} +
\chi_{00303} + \chi_{11211} + \chi_{30300}|^2 +\cr
& |\chi_{00203} + \chi_{11111} + \chi_{20013} + \chi_{30200}
+ \chi_{31002} + \chi_{01302} + \chi_{20310}|^2 +\cr
& |\chi_{01202} + \chi_{10104} + \chi_{20210} + \chi_{21012}
+ \chi_{40101} + \chi_{10401}|^2 +\cr
& |\chi_{01005} + \chi_{10301} + \chi_{11103} + \chi_{30111}
+ \chi_{50010} + \chi_{00500}|^2 +\cr
& |\chi_{00006} + \chi_{00400} + \chi_{02004} + \chi_{20202}
+ \chi_{40020} + \chi_{60000} + \chi_{00600}|^2
\cr\noalign{\medskip}}$$
$$\eqalign{
C(6,7) =& |\chi_{000000} + \chi_{010002} + \chi_{200010} + \chi_{000104}
+ \chi_{020020} + \chi_{101012} + \chi_{210101} + \chi_{401000} +\cr
&~~~~ \chi_{000007} + \chi_{002030} + \chi_{010122} + \chi_{030200}
+ \chi_{100024} + \chi_{111111} + \chi_{200203} + \chi_{221010} +\cr
&~~~~ \chi_{302002} + \chi_{420001} + \chi_{700000} + \chi_{000240}
+ \chi_{001042} + \chi_{012210} + \chi_{020302} + \chi_{042000} +\cr
&~~~~ \chi_{101221} + \chi_{122101} + \chi_{203020} + \chi_{240100}
+ \chi_{000070} + \chi_{002401} + \chi_{010420} + \chi_{024010} +\cr
&~~~~ \chi_{070000} + \chi_{104200} + \chi_{000700} + \chi_{007000}|^2 +\cr
&|\chi_{100001} + \chi_{001003} + \chi_{110011} + \chi_{300100}
+ \chi_{000015} + \chi_{011021} + \chi_{100113} + \chi_{120110} +\cr
&~~~~ \chi_{201102} + \chi_{311001} + \chi_{510000}
+ \chi_{001131} + \chi_{010033} + \chi_{021201} + \chi_{102120}
+ \chi_{110212} +\cr
&~~~~ \chi_{131100} + \chi_{212011} + \chi_{330010} + \chi_{000151}
+ \chi_{003300} + \chi_{011311} + \chi_{033001} + \chi_{100330} +\cr
&~~~~ \chi_{113110} + \chi_{151000} + \chi_{001510}
+ \chi_{015100}|^2 + 32 |\chi_{111111}|^2
\cr}\eqno(coninv)$$
\bigskip

\leftline{\bf A.4 $E_7$-type invariants}

$E_7$-type invariants are those that can be obtained acting with
an automorphism of the fusion rules of the extended algebra\refto{MS}.
It has been claimed that they only exist for $N=2,3,4,5,8,9,16$
$^{[\cite{Ver},\cite{AF}]}$. The $\widehat {SU}(3)_9$ and
the $\widehat {SU}(5)_5$ invariants were found in Refs. \[MS] and
\[SY2] respectively. The remaining ones were obtained in Refs.
\[Ver] and \[AF]. Below we list the $N=3,4,5$ invariants
that are relevant for our analysis. They are denoted by
$E(N-1,k)$.
$$\eqalign{
E(2,9) = &|\chi_{00} + \chi_{90} + \chi_{09}|^2 +
|\chi_{22} + \chi_{25} + \chi_{52}|^2
+ |\chi_{44} + \chi_{41} + \chi_{14}|^2 +
|\chi_{36} + \chi_{60} + \chi_{03}|^2 +\cr
& |\chi_{63} + \chi_{06} + \chi_{30}|^2 + 2|\chi_{33}|^2
+ \big [ (\chi_{11} + \chi_{17} + \chi_{71})\chi^*_{33} + c.c. \big ]
\cr\noalign{\medskip}
E(3,8) =&
|\chi_{000} + \chi_{800} + \chi_{080} + \chi_{008}|^2 +
|\chi_{400} + \chi_{440} + \chi_{044} + \chi_{004}|^2 +\cr
&  |\chi_{020} + \chi_{602} + \chi_{060} + \chi_{206}|^2 +
|\chi_{024} + \chi_{202} + \chi_{420} + \chi_{242}|^2 +\cr
&  |\chi_{032} + \chi_{303} + \chi_{230} + \chi_{323}|^2 +
|\chi_{311} + \chi_{331} + \chi_{133} + \chi_{113}|^2  +\cr
&  |\chi_{121} + \chi_{412} + \chi_{141} + \chi_{214}|^2 +
|\chi_{040} + \chi_{404}|^2 + 2|\chi_{222}|^2 +\cr
&  \big [ (\chi_{012} + \chi_{501}
 + \chi_{250} + \chi_{125})\chi_{222}^*
+ (\chi_{210} + \chi_{521} + \chi_{052} + \chi_{105} ) \chi_{222}^* +\cr
& (\chi_{101} + \chi_{610} + \chi_{161} + \chi_{016})
(\chi_{040} + \chi_{404})^*
 + c.c. \big ]
\cr\noalign{\medskip}}$$
$$\eqalign{
E(4,5) = &|\chi_{0000} + \chi_{5000} + \chi_{0500} + \chi_{0050} +
\chi_{0005}|^2 + |\chi_{0102} + \chi_{2010} + \chi_{0220} +
\chi_{1022} + \chi_{2201}|^2 +\cr
&  |\chi_{0110} + \chi_{3011} +
\chi_{0301} + \chi_{1030} + \chi_{1103}|^2
 + |\chi_{1200} + \chi_{2120} +
\chi_{0212} + \chi_{0021} + \chi_{2002}|^2  +\cr
& \big [(\chi_{1001} + \chi_{3100} +
\chi_{1310} + \chi_{0131} + \chi_{0013})\chi^*_{1111} + c.c \big ]
+ 4|\chi_{1111}|^2 \cr}\eqno(einv)$$

\vfill\eject

\references
\refis{Gepner89} D. Gepner, Phys. Lett. B222 (1989) 207.
\refis{G3} D. Gepner, ``String theory on Calabi-Yau manifolds: the three
generation case", Princeton Univ. preprint PUPT-88-0085 (1988).\par
\refis{GepnerW} D. Gepner, Nucl. Phys. B322 (1989) 65\par
\refis{Gepner87} D. Gepner, Nucl. Phys. B 296 (1988) 757;
Phys. Lett. B199 (1987) 380.\par
\refis{FIQS}A. Font, L.E. Ib\' a\~ nez, F. Quevedo and A. Sierra,
Nucl. Phys. B337 (1989) 119.\par
\refis{Fuchs}J. Fuchs, A. Klemm, C. Scheich and M. Schmidt,
Phys. Lett. B232 (1989) 232; Ann. Phys. 204 (1990) 1.\par
\refis{Kazama}Y. Kazama and H. Suzuki, Nucl. Phys. B321 (1989) 232.\par
\refis{Font}A. Font, L.E. Ib\'a\~nez and F. Quevedo,
Phys. Lett. B217 (1989) 271.\par
\refis{LVW} W. Lerche, C. Vafa and N.P. Warner,
Nucl. Phys. B324 (1989) 427.\par
\refis{GVW} B. Greene, C. Vafa and N.P. Warner,
Nucl. Phys. B324 (1989) 371.\par
\refis{Mar} E. Martinec, Phys. Lett. B217 (1989) 431. \par
\refis{schya} A.N. Schellekens and S. Yankielowicz,
Nucl. Phys. B330, (1990) 103.\par
\refis{Lutken} A. L\"utken and G.G. Ross,
 Phys. Lett. B213 (1988) 512. \par
\refis{Bailin} D. Bailin, D.C. Dunbar and A. Love, Int. J. Mod. Phys.
A6 (1991) 1659. \par
\refis{But1} E. Buturovic, Phys. Lett. B236 (1990) 277. \par
\refis{But2} E. Buturovic, Nucl. Phys. B352 (1991) 163. \par
\refis{VW} C. Vafa and N.P. Warner, Phys. Lett. B218 (1989) 51. \par
\refis{Vafa} C. Vafa, Mod. Phys. Lett. A4 (1989) 1169.\par
\refis{LS1} M. Lynker and R. Schimmrigk,
 Nucl. Phys. B339 (1990) 121.\par
\refis{LS2} M. Lynker and R. Schimmrigk, Phys. Lett. B253 (1991) 83.\par
\refis{Sche} A.N. Schellekens, ``Field identification fixed points in
$N=2$ coset theories", preprint CERN-TH.6055/91 (1991).\par
\refis{Schwarz} J.H. Schwarz, Int. J. Mod. Phys. A4 (1989) 2653. \par
\refis{Capelli} A. Capelli, C. Itzykson and J.B. Zuber,
Nucl. Phys. B280 (1987) 445.\par
\refis{Ber} D. Bernard, Nucl. Phys. B288 (1987) 628. \par
\refis{ALZ} D. Altschuler, J. Lacki and P. Zaugg,
Phys. Lett. B205 (1988) 281.\par
\refis{SY1} A.N. Schellekens and S. Yankielowicz,
Nucl. Phys. B327 (1989) 673; Phys. Lett. B227 (1989) 387.\par
\refis{SY2} A.N. Schellekens and S. Yankielowicz,
Nucl. Phys. B334 (1990) 67; Int. J. Mod. Phys. A5 (1990) 2903.\par
\refis{Ver} D. Verstegen, Nucl. Phys. B346 (1990) 349;
Comm. Math. Phys. 137 (1991) 567.\par
\refis{su3} P. Christe and F. Ravanini,
Int. J. Mod. Phys. A4 (1989) 897; \hfil\break
M. Bauer and C.Itzykson, Comm. Math. Phys. 127 (1990) 617.\par
\refis{dual} M. Walton, Nucl. Phys. B322 (1989) 775; \hfil\break
D. Altschuler, M. Bauer and C. Itzykson,
Comm. Math. Phys. 132 (1990) 349.\par
\refis{MS} G. Moore and N. Seiberg, Nucl. Phys. B313 (1988) 16.\par
\refis{BN}  P. Bouwknegt and W. Nahm, Phys. Lett. B184 (1987) 359.\par
\refis{emb} A.N. Schellekens and N.P. Warner. Phys. Rev. D34 (1986) 3092;
\hfil\break
F. Bais and P. Bouwknegt, Nucl. Phys. B279 (1987) 561;
\hfil\break
R. Arcuri, J. Gomes and D. Olive, Nucl. Phys. B285 (1987) 327. \par
\refis{AF} A. Font, ``Automorphism fixed points and exceptional
modular invariants", preprint FERMILAB-PUB-91/229-T.\par
\refis{Kac} V.G. Kac, {\it Infinite Dimensional Lie Algebras} (Cambridge
University Press, 1990). \par
\refis{Arnold} V.I. Arnold, S.M. Gusein-Zade and A.N. Varchenko, {\it
Singularities of Differentiable Maps}, vol.I, (Birkhauser, 1985).\par

\endreferences
\vfill\eject

\pageno 28
$$\vbox{\settabs 3\columns
\+ $N$ & $k$ & Invariant \cr
{\smallskip \hrule \bigskip}
\+ 3,5,7 & 0 mod $N$ & $D$ \cr
{\smallskip \hrule \bigskip}
\+ 2 & 0 mod 4 & $D$ \cr
\+  & 2 mod 4 & $F$ \cr
{\smallskip \hrule \bigskip}
\+ 4 & 0 mod 8 & $D$, $D_2$ \cr
\+  & 4 mod 8 & $F$, $D_2$ \cr
\+  & 2 mod 4 &  $D_2$ \cr
\+  & 1 mod 2 &  $G$ \cr
{\smallskip \hrule \bigskip}
\+ 6 & 0 mod 12 & $D$ \cr
\+  & 6 mod 12 & $F$ \cr
\+  & 0 mod 3 & $D_2$ \cr
\+  & 0 mod 4 & $D_3$ \cr
\+  & 2 mod 4 & $G$ \cr
{\smallskip \hrule }
}$$
\bigskip\medskip
\noindent
TABLE \ 2. Bernard invariants for $N \leq 7$.

\vfill\eject

$$\vbox{\settabs 4\columns
\+ $\bar G$ & $G$ & $\bar k$ & Invariant \cr
{\smallskip \hrule \bigskip}
\+ $SU(3)$ & $SO(8)$ & 3 & $D$ \cr
\+  & $SU(6)$ & 5 & $C(2,5)$ \cr
\+  & $E_6$ & 9 & $C(2,9)$ \cr
\+  & $E_7$ & 21 & $C(2,21)$ \cr
{\smallskip \hrule \bigskip}
\+ $SU(4)$ & $SU(6)$ & 2 & $D_2$ \cr
\+  & $SO(15)$ & 4 & $C(3,4)$ \cr
\+  & $SU(10)$ & 6 & $C(3,6)$ \cr
\+  & $SO(20)$ & 8 & $C(3,8)$ \cr
{\smallskip \hrule \bigskip}
\+ $SU(5)$ & $SU(10)$ & 3 & $C(4,3)$ \cr
\+  & $SO(24)$ & 5 & $C(4,5)$ \cr
\+  & $SU(15)$ & 7 & $C(4,7)$ \cr
{\smallskip \hrule \bigskip}
\+ $SU(6)$ & $SU(15)$ & 4 & $C(5,4)$ \cr
\+  & $SO(35)$ & 6 & $C(5,6)$ \cr
\+  & $Sp(20)$ & 6 & $\tilde C(5,6)$ \cr
\+  & $SU(21)$ & 8 & $C(5,8)$ \cr
{\smallskip \hrule \bigskip}
\+ $SU(7)$ & $SU(21)$ & 5 & $C(6,5)$ \cr
\+  & $SO(48)$ & 7 & $C(6,7)$ \cr
\+  & $SU(28)$ & 9 & $C(6,9)$ \cr
{\smallskip \hrule }
}$$
\bigskip
\noindent
TABLE \ 3. Conformal embeddings of $\sua$, $N=3, \cdots, 7$ into simple
groups.

\vfill\eject\end

\pageno 23
\advance\vsize by \baselineskip

\centerline{\bf TABLE 1}
\bigskip
\halign{\indent#\hfil&\quad#\hfil&\quad\hfil#\hfil&\quad\hfil#\hfil&
\quad\hfil#\hfil&\quad\hfil#\hfil&\quad\hfil#\hfil&\quad\hfil#\hfil&
\quad#\hfil&\quad#\hfil&\quad#\hfil&\quad#\hfil\cr
 & & model& & & & & &invariant&N27&N$\overline {27}$&$N_{gen}$\cr
\noalign{\hrule}
 1:&(6,7)& & & & & & &D&~62&~~2&~60\cr
 & & & & & & & &C&~15&~15&~~0\cr
\noalign{\hrule}
2:&(5,9)& & & & & & &$D_2$&~66&~~3&~63\cr
\noalign{\hrule}
3:&(4,15)& & & & & & &D&~99&~~3&~96\cr
\noalign{\hrule}
4:&(4,5)&(2,3)& & & & & &AD, DD&~41&~~5&~36\cr
  & & & & & & & &CD, ED&~31&~~7&~24\cr
\noalign{\hrule}
5:&(5,6)&2& & & & & &FA, $D_2$A, GA&~61&~~1&~60\cr
  & & & & & & & &CA&~15&~15&~~0\cr
  & & & & & & & &$\tilde C$A&~27&~~3&~24\cr
\noalign{\hrule}
7:&(4,7)&4& & & & & &CA&~24&~12&~12\cr
\noalign{\hrule}
8:&(4,10)&1& & & & & &DA&~66&~~3&~63\cr
\noalign{\hrule}
9:&(4,4)&1&16& & & & &AAD&~76&~~4&~72\cr
  & & & & & & & &AAE&112&~~4&108\cr
\noalign{\hrule}
10:&(4,5)&1&4&  & & & &DAA&~41&~~5&~36\cr
  & & & & & & & &CAA, EAA&~31&~~7&~24\cr
\noalign{\hrule}
11:&(4,5)&2&2&  & & & &DAA&~41&~~5&~36\cr
  & & & & & & & &CAA, EAA&~31&~~7&~24\cr
\noalign{\hrule}
12:&(4,7)&1&1& & & & &CAA&~22&~10&~12\cr
\noalign{\hrule}
13:&(4,5)&1&1&1& & & &DAAA, CAAA, EAAAA&~21&~21&~~0\cr
\noalign{\hrule}
14:&(3,4)&(3,4)& &  & & & &F$D_2$&~43&~~3&~40\cr
  & & & & & & & &FF, $D_2D_2$&~59&~~3&~56\cr
 & & &  & & & & &FC, $D_2$C&~31&~~7&~24\cr
  & & & & & & & &CC&~15&~15&~~0\cr
\noalign{\hrule}
15:&(3,3)&(2,18)& & & & & &AD, GA, GD&~92&~~8&~84\cr
\noalign{\hrule}
16:&(3,4)&(2,9)& & & & & &$D_2$D&~55&~~7&~48\cr
  & & & & & & & &FC, $D_2$C&~31&~~7&~24\cr
  & & & & & & & &$D_2$E&~45&~~5&~40\cr
  & & & & & & & &FD&~35&~11&~24\cr
  & & & & & & & &FE&~37&~~5&~32\cr
  & & & & & & & &CD&~27&~15&~12\cr
  & & & & & & & &CC&~15&~15&~~0\cr
  & & & & & & & &CE&~23&~11&~12\cr
\noalign{\hrule}
17:&(3,5)&(2,6)& & & & & &GA&~86&~~2&~84\cr
 & & & & & & & &GD&~44&~~8&~36\cr
\noalign{\hrule}
18:&(3,8)&(2,3)& & & & & &$D_2$D&~45&~~9&~36\cr
  & & & & & & & &DD, CD, ED&~21&~21&~~0\cr
\noalign{\hrule}
20:&(3,3)&(2,3)&5& & & & &ADA, GDA&~21&~21&~~0\cr
\noalign{\hrule}
21:&(3,4)&(2,3)&2& & & & &FDA, $D_2$DA&~51&~~3&~48\cr
  & & & & & & & &CDA&~31&~~7&~24\cr
\noalign{\hrule}
22:&(3,5)&(2,3)&1& & & & &GDA&~21&~21&~~0\cr
\noalign{\hrule}
23:&(3,5)&1&1&1&1& & &GAAAA&~62&~~2&~60\cr
\noalign{\hrule}
24:&(3,4)&1&1&1&2& & &FAAAA, $D_2$AAAA, CAAAA&~21&~21&~~0\cr
\noalign{\hrule}
25:&(3,3)&1&1&1&5& & &GAAAA&~21&~21&~~0\cr
\noalign{\hrule}
26:&(3,8)&1&1&1& & & &DAAA, $D_2$AAA&~62&~~2&~60\cr
 & & & & & & & &CAAA&~29&~~5&~24\cr
 & & & & & & & &EAAA&~40&~~4&~36\cr
\noalign{\hrule}
27:&(3,5)&1&1&4& & & &GAAA&~62&~~2&~60\cr
\noalign{\hrule}
28:&(3,4)&1&1&10& & & &FAAA, $D_2$AAF&~33&~~9&~24\cr
  & & & & & & & &FAAF, $D_2$AAA&~37&~13&~24\cr
  & & & & & & & &CAAA, CAAF&~19&~19&~~0\cr
\noalign{\hrule}
%\vadjust{\eject}
29:&(3,5)&1&2&2& & & &GAAA&~21&~21&~~0\cr
\noalign{\hrule}
30:&(3,4)&1&2&4& & & &$D_2$AAA, FAAA&~51&~~3&~48\cr
 & & & & & & & &CAAA&~31&~~7&~24\cr
\noalign{\hrule}
31:&(3,3)&1&4&5& & & &GAAA&~21&~21&~~0\cr
\noalign{\hrule}
32:&(3,4)&2&2&2& & & &FAAA&~49&~~1&~48\cr
  & & & & & & & &$D_2$AAA&~73&~~1&~72\cr
\vadjust{\eject}
  & & & & & & & &CAAA&~39&~~3&~36\cr
& &model& & & & & &invariant&N27&N$\overline {27}$&$N_{gen}$\cr
\noalign{\hrule}
33:&(3,3)&2&2&5& & & &GAAA&~21&~21&~~0\cr
\noalign{\hrule}
34:&(3,14)&1&1& & & & &$D_2$AA&~79&~~7&~72\cr
\noalign{\hrule}
35:&(3,8)&1&4& & & & &DAA&~62&~~2&~60\cr
 & & & & & & & &$D_2$AA&~74&~~2&~72\cr
 & & & & & & & &CAA&~29&~~5&~24\cr
 & & & & & & & &EAA&~40&~~4&~36\cr
\noalign{\hrule}
36:&(3,6)&1&13& & & & &$D_2$AA&~55&~~7&~48\cr
   & & & & & & & &CAA&~35&~11&~24\cr
\noalign{\hrule}
37:&(3,8)&2&2& & & & &DAA, CAA, EAA&~21&~21&~~0\cr
  & & & & & & & &$D_2$AA&~45&~~9&~36\cr
\noalign{\hrule}
38:&(3,5)&2&10& & & & &GAA, GAF&~44&~~8&~36\cr
\noalign{\hrule}
39:&(3,6)&3&3& & & & &$D_2$AA&~75&~~3&~72\cr
  & & & & & & & &CAA&~49&~~5&~44\cr
\noalign{\hrule}
40:&(3,4)&3&18& & & & &$D_2$AF, FAA&~27&~19&~~8\cr
  & & & & & & & &FAF, $D_2$AA&~51&~11&~40\cr
  & & & & & & & &CAA, CAF&~23&~23&~~0\cr
\noalign{\hrule}
41:&(3,5)&4&4& & & & &GAA&~53&~~5&~48\cr
\noalign{\hrule}
42:&(3,4)&4&10& & & & &$D_2$AF, FAA&~35&~11&~24\cr
  & & & & & & & &$D_2$AA, FAF&~55&~~7&~48\cr
  & & & & & & & &CAA, CAF&~27&~15&~12\cr
\noalign{\hrule}
43:&(3,4)&6&6& & & & &FAA, FFF&~41&~~9&~32\cr
  & & & & & & & &FAF&~43&~~3&~40\cr
  & & & & & & & &$D_2$AA, $D_2$FF&~69&~~5&~64\cr
  & & & & & & & &$D_2$AF&~59&~~3&~56\cr
  & & & & & & & &CAA, CFF&~35&~11&~24\cr
  & & & & & & & &CAF&~31&~~7&~24\cr
\noalign{\hrule}
44:&(3,3)&6&54& & & & &AAF, AFA, GAF, GFA&~54&~14&~40\cr
  & & & & & & & &AFF, GAA, GFF&~76&~20&~56\cr
\noalign{\hrule}
45:&(3,3)&12&12& & & & &AAD, ADD, GAD, GDD&~90&~~2&~88\cr
  & & & & & & & &GAA&121&~~5&116\cr
\noalign{\hrule}
46:&(3,32)&1& & & & & &DA, $D_2$A&185&~~5&180\cr
\noalign{\hrule}
47:&(3,20)&2& & & & & &FA, $D_2$A&114&~~6&108\cr
\noalign{\hrule}
48:&(3,16)&3& & & & & &DA, $D_2$A&~99&~~3&~96\cr
\noalign{\hrule}
49:&(3,14)&4& & & & & &$D_2$A&~79&~~7&~72\cr
\noalign{\hrule}
50:&(3,12)&6& & & & & &AF, FA, $D_2$F&~73&~~5&~68\cr
  & & & & & & & &FF, $D_2$A&103&~~3&100\cr
\noalign{\hrule}
51:&(3,11)&8& & & & & &GA, GD&~77&~~5&~72\cr
\noalign{\hrule}
52:&(3,10)&12& & & & & &AD, $D_2$A, $D_2$D&~85&~~5&~80\cr
\noalign{\hrule}
53:&(3,9)&24& & & & & &AD, GA, GD&117&~~5&112\cr
\noalign{\hrule}
54:&(2,3)&(2,3)&(2,3)& & & & &DDD&~51&~~3&~48\cr
\noalign{\hrule}
56:&(2,4)&(2,39)& & & & & &AD&155&~11&144\cr
\noalign{\hrule}
57:&(2,5)&(2,21)& & & & & &CD&~64&~16&~48\cr
  & & & & & & & &CC&~19&~19&~~0\cr
\noalign{\hrule}
58:&(2,6)&(2,15)& & & & & &AD, DD&~50&~14&~36\cr
\noalign{\hrule}
60:&(2,9)&(2,9)& & & & & &DD&~52&~10&~42\cr
  & & & & & & & &DC&~27&~15&~12\cr
  & & & & & & & &DE&~41&~11&~30\cr
  & & & & & & & &CC&~15&~15&~~0\cr
  & & & & & & & &CE&~23&~11&~12\cr
  & & & & & & & &EE&~34&~~8&~26\cr
\noalign{\hrule}
62:&(2,3)&(2,3)&1&1&1& & &DDAAA&~21&~21&~~0\cr
\noalign{\hrule}
63:&(2,3)&(2,6)&1&1& & & &DAAA, DDAA&~21&~21&~~0\cr
\noalign{\hrule}
64:&(2,3)&(2,3)&1&4& & & &DDAA&~51&~~3&~48\cr
\noalign{\hrule}
65:&(2,3)&(2,3)&2&2& & & &DDAA&~51&~~3&~48\cr
\noalign{\hrule}
76:&(2,3)&(2,4)&12& & & & &DAA&~38&~20&~18\cr
\vadjust{\eject}
  & & & & & & & &DAD&~21&~21&~~0\cr}

\halign{\indent#\hfil&\quad#\hfil&\quad\hfil#\hfil&\quad\hfil#\hfil&
\quad\hfil#\hfil&\quad\hfil#\hfil&\quad\hfil#\hfil&\quad\hfil#\hfil&
\quad#\hfil&\quad#\hfil&\quad#\hfil&\quad#\hfil\cr
 & & model& & & & & &invariant&N27&N$\overline {27}$&$N_{gen}$\cr
\noalign{\hrule}
77:&(2,3)&(2,5)&6& & & & &DCA, DCF&~35&~11&~24\cr
\noalign{\hrule}
78:&(2,3)&(2,6)&4& & & & &DAA, DDA&~23&~23&~~0\cr
\noalign{\hrule}
80:&(2,3)&(2,9)&2& & & & &DDA&~45&~~9&~36\cr
  & & & & & & & &DCA&~31&~~7&~24\cr
  & & & & & & & &DEA&~41&~~5&~36\cr
\noalign{\hrule}
81:&(2,3)&(2,15)&1& & & & &DDA&~43&~19&~24\cr
\noalign{\hrule}
83:&(2,5)&(2,5)&2& & & & &CCA&~33&~~9&~24\cr
\noalign{\hrule}
84:&(2,6)&(2,6)&1& & & & &ADA, DDA&~44&~~8&~36\cr
\noalign{\hrule}
85:&(2,3)&1&1&1&1&1&1&DAAAAAA&~21&~21&~~0\cr
\noalign{\hrule}
86:&(2,6)&1&1&1&1&1& &DAAAAA&~62&~~2&~60\cr
\noalign{\hrule}
87:&(2,3)&1&1&1&1&4& &DAAAAA&~35&~11&~24\cr
\noalign{\hrule}
88:&(2,3)&1&1&1&2&2& &DAAAAA&~21&~21&~~0\cr
\noalign{\hrule}
91:&(2,15)&1&1&1&1& & &DAAAA&~68&~~8&~60\cr
\noalign{\hrule}
92:&(2,9)&1&1&1&2& & &DAAAA, CAAAA, EAAAA&~21&~21&~~0\cr
\noalign{\hrule}
94:&(2,6)&1&1&1&4& & &DAAAA&~62&~~2&~60\cr
\noalign{\hrule}
95:&(2,5)&1&1&1&6& & &CAAAA, CAAAF&~21&~21&~~0\cr
\noalign{\hrule}
96:&(2,4)&1&1&1&12& & &AAAAD&~21&~21&~~0\cr
\noalign{\hrule}
97:&(2,6)&1&1&2& 2& & &DAAAA&~21&~21&~~0\cr
\noalign{\hrule}
98:&(2,3)&1&1&2&10& & &DAAAA, DAAAF&~35&~11&~24\cr
\noalign{\hrule}
99:&(2,3)&1&1&4&4& & &DAAAA&~51&~~3&~48\cr
\noalign{\hrule}
100:&(2,3)&1&2&2&4& & &DAAAA&~51&~~3&~48\cr
\noalign{\hrule}
101:&(2,3)&2&2&2&2& & &DAAAA&~61&~~1&~60\cr
\noalign{\hrule}
107:&(2,33)&1&1&2& & & &DAAA&100&~16&~84\cr
\noalign{\hrule}
108:&(2,15)&1&1&4& & & &DAAA&~79&~~7&~72\cr
\noalign{\hrule}
109:&(2,9)&1&1&10& & & &DAAA&~34&~22&~12\cr
 & & & & & & & &DAAF, EAAA&~28&~16&~12\cr
 & & & & & & & &CAAA, CAAF&~19&~19&~~0 \cr
 & & & & & & & &EAAF&~26&~14&~12\cr
\noalign{\hrule}
111:&(2,15)&1&2&2& & & &DAAA&~43&~19&~24\cr
\noalign{\hrule}
112:&(2,9)&1&2&4& & & &DAAA&~45&~~9&~36\cr
 & & & & & & & &CAAA&~31&~~7&~24\cr
 & & & & & & & &EAAA&~41&~~5&~36\cr
\noalign{\hrule}
113:&(2,6)&1&2&10& & & &AAAF, DAAA, DAAF&~44&~~8&~36\cr
\noalign{\hrule}
114:&(2,5)&1&2&22& & & &CAAA, CAAF&~37&~13&~24\cr
\noalign{\hrule}
116:&(2,6)&1&4&4& & & &DAAA&~53&~~5&~48\cr
\noalign{\hrule}
117:&(2,5)&1&4&6& & & &CAAA, CAAF&~35&~11&~24\cr
\noalign{\hrule}
118:&(2,4)&1&4&12& & & &AAAD&~21&~21&~~0\cr
\noalign{\hrule}
119:&(2,3)&1&5&40& & & &DAAA&~35&~35&~~0\cr
%\vadjust{\eject}
 & & & & & & & &DAAD&~21&~21&~~0\cr
%\halign{\indent#\hfil&\quad#\hfil&\quad\hfil#\hfil&\quad\hfil#\hfil&
%\quad\hfil#\hfil&\quad\hfil#\hfil&\quad\hfil#\hfil&\quad\hfil#\hfil&
%\quad#\hfil&\quad#\hfil&\quad#\hfil&\quad#\hfil\cr
% & & model& & & & & &invariant&N27&N$\overline {27}$&$N_{gen}$\cr
\noalign{\hrule}
120:&(2,3)&1&6&22& & & &DAAA, DAFF&~43&~19&~24\cr
 & & & & & & & &DAAF, DAFA&~35&~11&~24\cr
\noalign{\hrule}
121:&(2,3)&1&7&16& & & &DAAA&~43&~19&~24\cr
 & & & & & & & &DAAD, DAAE&~21&~21&~~0\cr
\noalign{\hrule}
122:&(2,3)&1&8&13& & & &DAAA&~27&~27&~~0\cr
 & & & & & & & &DADA&~21&~21&~~0\cr
\noalign{\hrule}
123:&(2,3)&1&10&10& & & &DAAA, DAFF&~59&~11&~48\cr
 & & & & & & & &DAAF&~51&~~3&~48\cr
\noalign{\hrule}
129:&(2,9)&2&2&2& & & &DAAA&~69&~~3&~66\cr
 & & & & & & & &CAAA&~39&~~3&~36\cr
 & & & & & & & &EAAA&~56&~~2&~54\cr
\noalign{\hrule}
131:&(2,6)&2&2&4& & & &DAAA&~23&~23&~~0\cr
\noalign{\hrule}
132:&(2,5)&2&2&6& & & &CAAA, CAAF&~52&~~4&~48\cr
\noalign{\hrule}
133:&(2,4)&2&2&12& & & &AAAD&~21&~21&~~0\cr
\noalign{\hrule}
134:&(2,3)&2&3&18& & & &DAAA, DAAF&~39&~15&~24\cr
\noalign{\hrule}
135:&(2,3)&2&4&10& & & &DAAA, DAAF&~45&~~9&~36\cr
\noalign{\hrule}
137:&(2,3)&2&6&6& & & &DAAA,DAFF&~55&~~7&~48\cr
\vadjust{\eject}
 & & & & & & & &DAAF&~51&~~3&~48\cr
 & & model& & & & & &invariant&N27&N$\overline {27}$&$N_{gen}$\cr
\noalign{\hrule}
141:&(2,3)&3&3&8& & & &DAAA&~39&~15&~24\cr
 & & & & & & & &DAAD&~21&~21&~~0\cr
\noalign{\hrule}
144:&(2,3)&4&4&4& & & &DAAA&~60&~~6&~54\cr
\noalign{\hrule}
146:&(2,123)&1&5& & & & &DAA&377&~17&360\cr
\noalign{\hrule}
147:&(2,69)&1&6& & & & &DAA&242&~14&228\cr
  & & & & & & & &DAF&176&~20&156\cr
\noalign{\hrule}
148:&(2,51)&1&7& & & & &DAA&214&~10&204\cr
\noalign{\hrule}
149:&(2,42)&1&8& & & & &AAD, DAA, DAD&125&~17&108\cr
\noalign{\hrule}
150:&(2,33)&1&10& & & & &DAA&185&~~5&180\cr
  & & & & & & & &DAF&121&~13&108\cr
\noalign{\hrule}
151:&(2,27)&1&13& & & & &DAA&~87&~15&~72\cr
\noalign{\hrule}
152:&(2,24)&1&16& & & & &AAD, DAA, DAD&122&~~8&114\cr
  & & & & & & & &AAE, DAE&~83&~14&~69\cr
\noalign{\hrule}
153:&(2,21)&1&22& & & & &DAA&~96&~12&~84\cr
  & & & & & & & &DAF&~46&~34&~16\cr
  & & & & & & & &CAA, CAF&~23&~23&~~0\cr
\noalign{\hrule}
155:&(2,18)&1&40& & & & &AAD&198&~~6&192\cr
  & & & & & & & &DAA, DAD&~65&~29&~36\cr
\noalign{\hrule}
157:&(2,16)&1&112& & & & &AAD&152&~26&126\cr
\noalign{\hrule}
158:&(2,57)&2&3& & & & &DAA&173&~17&156\cr
\noalign{\hrule}
%\vadjust{\eject}
159:&(2,33)&2&4& & & & &DAA&126&~12&114\cr
\noalign{\hrule}
161:&(2,21)&2&6& & & & &DAA&114&~~6&108\cr
  & & & & & & & &DAF&~83&~11&~72\cr
  & & & & & & & &CAA, CAF&~33&~~9&~24\cr
\noalign{\hrule}
163:&(2,15)&2&10& & & & &DAA, DAF&~70&~10&~60\cr
\noalign{\hrule}
165:&(2,12)&2&18& & & & &AAF, DAA, DAF&~44&~26&~18\cr
\noalign{\hrule}
167:&(2,10)&2&50& & & & &AAF&~95&~23&~72\cr
\noalign{\hrule}
168:&(2,27)&3&3& & & & &DAA&117&~~9&108\cr
\noalign{\hrule}
169:&(2,12)&3&8& & & & &AAD, DAA, DAD&~77&~~5&~72\cr
\noalign{\hrule}
170:&(2,9)&3&18& & & & &DAA&~49&~13&~36\cr
  & & & & & & & &DAF&~21&~33&~12\cr
  & & & & & & & &CAA, CAF&~23&~23&~~0\cr
  & & & & & & & &EAA&~37&~17&~20\cr
  & & & & & & & &EAF&~25&~21&~4\cr
\noalign{\hrule}
171:&(2,15)&4&4& & & & &DAA&~99&~~3&~96\cr
\noalign{\hrule}
172:&(2,9)&4&10& & & & &DAA&~52&~10&~42\cr
  & & & & & & & &DAF&~28&~22&~~6\cr
  & & & & & & & &CAA, CAF&~27&~15&~12\cr
  & & & & & & & &EAA&~41&~11&~30\cr
  & & & & & & & &EAF&~31&~13&~18\cr
\noalign{\hrule}
175:&(2,6)&5&40& & & & &AAD, DAA, DAD&~35&~35&~~0\cr
\noalign{\hrule}
176:&(2,9)&6&6& & & & &DAA, DFF&~66&~~6&~60\cr
  & & & & & & & &DAF&~55&~~7&~48\cr
  & & & & & & & &CAA, CFF&~35&~11&~24\cr
  & & & & & & & &CAF&~31&~~7&~24\cr
  & & & & & & & &EAA, EFF&~52&~~8&~44\cr
  & & & & & & & &EAF&~45&~~5&~40\cr
\noalign{\hrule}
177:&(2,6)&6&22& & & & &AAF, AFA, DFA, DAF&~26&~38&~12\cr
  & & & & & & & &AFF, DAA, DFF&~38&~26&~12\cr
\noalign{\hrule}
178:&(2,6)&7&16& & & & &AAD&123&~~3&120\cr
  & & & & & & & &AAE&~86&~~2&~84\cr
  & & & & & & & &DAA, DAD&~50&~14&~36\cr
  & & & & & & & &DAE&~44&~~8&~36\cr
\noalign{\hrule}
179:&(2,5)&7&70& & & & &CAA, CAF&~47&~47&~~0\cr
\noalign{\hrule}
181:&(2,6)&8&13& & & & &ADA, DAA, FDA&~29&~29&~~0\cr
\noalign{\hrule}
182:&(2,5)&8&38& & & & &CAA, CAF&~52&~28&~24\cr
  & & & & & & & &CDA, CDF&~33&~33&~~0\cr
\noalign{\hrule}
183:&(2,6)&10&10& & & & &AAF, DAF&~41&~17&~24\cr
\vadjust{\eject}
  & & & & & & & &AFF, DAA, DFF&~59&~11&~48\cr
 & &model& & & & & &invariant&N27&N$\overline {27}$&$N_{gen}$\cr
\noalign{\hrule}
184:&(2,5)&10&22& & & & &CAA, CAF&~64&~16&~48\cr
 & & & & & & & &CFA, CFF&~40&~16&~24\cr
\noalign{\hrule}
185:&(2,4)&13&208& & & & &AAD&104&~62&~42\cr
\noalign{\hrule}
186:&(2,5)&14&14& & & & &CAA, CFF&~83&~11&~72\cr
  & & & & & & & &CAF&~57&~~9&~48\cr
\noalign{\hrule}
187:&(2,4)&14&110& & & & &AAF, AFA&~65&~65&~~0\cr
  & & & & & & & &AFF&110&~38&~72\cr
\noalign{\hrule}
188:&(2,4)&16&61& & & & &ADA&~71&~41&~30\cr
  & & & & & & & &AEA&~50&~29&~21\cr
\noalign{\hrule}
189:&(2,4)&19&40& & & & &AAD&155&~11&144\cr
\noalign{\hrule}
190:&(2,4)&26&26& & & & &AAF&116&~14&102\cr
  & & & & & & & &AFF&194&~~8&186\cr}

\end